\documentclass[12pt]{book}

\usepackage[dvips]{graphicx,color}
\usepackage{makeidx,tsukuba}

\makeauthorindex
\makeindex

\begin{document}

\BookTitle{\itshape The 28th International Cosmic Ray Conference}
\CopyRight{\copyright 2003 by Universal Academy Press, Inc.}
\pagenumbering{arabic}

\chapter{
Testing Scenarios of Lorentz Symmetry Violation \\ 
Generated at the 
Planck Scale}

\author{%
%
%
Luis Gonzalez-Mestres\\
{\it LAPP, CNRS-IN2P3, B.P. 110 , 74941 Annecy-le-Vieux Cedex}\\
}

\section*{Abstract}
Using new theoretical tools, which allow to better understand ultra-high 
energy (UHE) dynamics, several patterns of Lorentz symmetry violation (LSV) 
are studied and compared with experiment. It is claimed that quadratically 
deformed relativistic kinematics (QDRK), where the parameter driving LSV 
varies like the square of the energy scale, remains the best suited pattern 
to describe LSV generated at the Planck scale. Implications of existing data 
are discussed and prospects are presented having in mind next-generation experiments.

\section{Deformed Relativistic Kinematics}

We have proposed [4-7] several LSV patterns 
in the vacuum rest frame (VRF) producing 
deformed Lorentz symmetries (DLS) and 
deformed relativistic kinematics (DRK).   
In particular, QDRK [7-16] is 
characterized by:
\equation
E~=~~(2\pi )^{-1}~h~c~a^{-1}~e~(k~a)
\endequation
\noindent
$E$ being the energy, 
$h$ the Planck constant, $a$ the fundamental length, $c$ the speed of light, $k$ the wave vector
and
$[e~(k~a)]^2$ a convex
function of $(k~a)^2$ .
For $k~a~\ll ~1$~: 
\begin{eqnarray}
e~(k~a) & \simeq & [(k~a)^2~-~\alpha ~(k~a)^4~
+~(2\pi ~a)^2~h^{-2}~m^2~c^2]^{1/2}
\end{eqnarray}
\noindent
$\alpha $ being a model-dependent constant that may be in the range $0.1~-~0.01$ for
full-strength violation of Lorentz symmetry at the fundamental length scale,
and {\it m} the mass of the particle. For momentum $p~\gg ~mc$ , we get:
\begin{eqnarray}
E & \simeq & p~c~+~m^2~c^3~(2~p)^{-1}~
-~p~c~\alpha ~(k~a)^2/2~~~~~
\end{eqnarray}

LDRK (linearly deformed relativistic kinematics) 
is another fashionable model (see [1,2] and [14,15,17,18]) and corresponds to:
\begin{eqnarray}
e~(k~a) ~ \simeq ~ [(k~a)^2~-~\beta ~(k~a)^3~
+~(2\pi ~a)^2~h^{-2}~m^2~c^2]^{1/2}
\end{eqnarray}
\noindent
$\beta $ being a model-dependent constant.
For $p~\gg ~mc$ :
\begin{eqnarray}
E ~ \simeq ~ p~c~+~m^2~c^3~(2~p)^{-1}~
-~p~c~\beta ~(k~a)/2~~~~~
\end{eqnarray}

Consistency requires that, for large bodies, $\alpha ~\propto  ~m^{-2}$ and
$\beta  ~\propto  ~m^{-1}$ . More generally, DRK models can be built where
for $k~a~\ll ~1$ and $p~\gg ~mc$ the deformation term in the expression of $E$
varies like $-~ \kappa ~p~(k~a)^{\delta }$ , $\delta $ being 
an arbitrary positive real number and $\kappa $ a constant. Then, consistency
requires $\kappa ~\propto  ~m^{-\delta }$ for large bodies. Other models can be generated from 
the mixing of superbradyons with "ordinary" particles (see [14] and references 
therein, as well as updated discussions in [17,18]). In another paper submitted 
to this conference, we discuss some potentialities of superdradyon mixing in
DRK phenomenology. 

As shown in [17,18], both QDRK and LDRK can be obtained from formal
Lorentz symmetric 
patterns using suitable transformations. The approach folowed there is an extension
of that proposed by Magueijo and Smolin [21,22] within the Doubly Special Relativity scheme 
(see, f.i., [1,2] and references therein). The pattern of [17,18] allows to
make the transformation operators formally unitary and 
to use 
extra dimensions in the dynamical description of 4-dimensional DRK. It also makes possible 
to build extensions of the Kirzhnits-Chechin (KCh) model [19,20] generalizing its 
Finsler space approach and setting a natural bridge with QDRK. 
Such as initially formulated, the KCh model was unable to really
account for the absence of the Greisen-Zatsepin-Kuzmin (GZK) cutoff and, although it satisfied
the requirement $\alpha ~\propto  ~m^{-2}$ for large bodies when set in a form close to QDRK,
it presented the same feature for elementary particles which is an obvious drawback [18].   
Our approach allows to build new models similar in their formal structure 
to the KCh model but able to
reproduce the QDRK pattern.

A characteristic feature of DRK models is that, above some critical energy scale 
and even for a very small LSV, the 
properties of the ultra-high energy particles (UHEPs) drastically change. This is true not
only for their apparent
interaction properties, but also for their internal structure. In previous 
papers [14,15], we had already pointed out that the interaction properties of a ultra-high energy
cosmic ray (UHECR) particle 
in the LDRK pattern were expected
to be modified too early, as energy increases, for consistent phenomenology. The same conclusion 
can be reached by looking at the internal structure of the UHEP using the operator formalism
[17,18]. 
The basic 
mechanism can be illustrated from a simple soliton model developed in
[7-10].  Assume that, in the VRF, a wave function in a relativistic dynamical model is perturbed 
by nonlocal corrections with distance scale $a$, so that the initial Lorentz 
symmetry is violated and we get:
\equation
\phi ~(x~\pm~a)~~-~~\phi ~(x)~~~=
~~~\Sigma _{n=0}^{\infty }~\phi ^{(n)}~(x)~~(\pm a)^n
~(n!)^{-1}
\endequation
where $\phi ^{(n)}~=~d^n\phi /dx^n$ , $\phi ^0$ is the 
initial solution with Lorentz symmetry and
$\phi ~(x)$ is the fixed-time
wave function. 
Then, if the model has solitons, the natural
dimensionless parameter to describe LSV and to be compared with $\gamma _R^{-2}$ 
($\gamma _R$ is the unperturbed relativistic Lorentz factor), will be 
$\xi _2 ~=~\alpha ~(a~\gamma _R)^2~\Delta ^{-2}$ for QDRK and
$\xi _1~=~\beta ~a~\gamma _R~\Delta ^{-1}$ for LDRK, where $\Delta $ is the characteristic
size scale of the unperturbed relativistic soliton at rest [9,10,17,18]. 
When $\xi _1$ ($\xi _2$) becomes larger 
than $\gamma _R^{-2}$ , the 
effective size scale of the UHE soliton changes. It can be readily checked that
the change in size  
of the soliton induced by LSV 
occurs at lower energies in the case of LDRK, leading to phenomenological problems
[17,18] .

\section{QDRK}

The deformation approximated by
$\Delta ~E~=~-~p~c~\alpha ~(k~a)^2/2$ in the right-hand
side of (3) has important consequences for UHECR. 
At energies above
$E_{trans}~
\approx ~\pi ^{-1/2}~ h^{1/2}~(2~\alpha )^{-1/4}~a^{-1/2}~m^{1/2}~c^{3/2}$,
the deformation $\Delta ~E$
dominates over
the mass term $m^2~c^3~(2~p)^{-1}$ and modifies all
kinematical balances. For a proton, taking $a$ to be the Planck length,
a typical scale for $E_{trans}$ is $\sim ~ 10^{19}~eV$ . 
Furthermore, there is a sharp fall of partial and total cross-sections
for cosmic-ray energies above
$E_{lim} ~\approx ~(2~\pi )^{-2/3}~(E_T~a^{-2}~ \alpha ^{-1}~h^2~c^2)^{1/3}$,
where $E_T$ is the target energy. For microwave background photons and $\alpha~\approx ~0.1$ ,
$E_{lim} ~\approx ~10^{19}~eV$ .
Such figures naturally
allow to explain the absence of GZK cutoff. The change in internal structure of the proton is 
expected to occur at a similar energy scale. This is compatible with experimental data
and leads to interesting predictions that can be checked by future experiments [14-18].
Using the operator formalism and going to the rest frame of the UHECR, it is 
possible to monitor 
the evolution of its internal structure, not only for the standard QDRK model but also for 
more general patterns. Suitable techniques can be to go to the formally 
symmetric reference system, perform the Lorentz transformation and go back to the 
physical reference system,
or to directly write the deformation as a spontaneous LSV in the physical reference system.
The analysis of 
the internal structure of a UHEP in its rest frame suggests in turn more precise tests 
requiring a careful study of the first interactions of UHECR in the atmosphere. This
will be, if feasible, a difficult experimental task but it may be possible with next 
generation experiments [17,18].  

\section{LDRK}

LDRK was discarded in our 1997
and subsequent DRK papers, but is often proposed 
(see [1,2] and references therein) for cosmic-ray, $TeV$ gamma-ray
and gravitational-wave phenomenology.
The deformation
yields effects  
at comparatively low energies, with 
$E_{trans}~
\approx ~\pi ^{-1/3}~ h^{1/3}~(2~\beta )^{-1/3}~a^{-1/3}~m^{2/3}~c^{5/3}$ and
$E_{lim} ~\approx ~(2~\pi )^{-1/2}~(E_T~a^{-1} \beta ^{-1}~h~c)^{1/2}$ .
The $E_{trans}$ scale, for a UHE 
proton, is now $\sim ~ 10^{16}~eV$ and similarly
for the change of its internal structure. For atmospheric targets, $E_{lim}$ 
falls naturally in the $\sim ~ 10^{19}~eV$ region if $a$ is the Planck length.
No existing phenomenological analysis or experimental data tend to confirm such predictions.
Although LDRK is often used
to explain possible anomalies of data at comparatively low energy ($TeV$ scale), 
the basic difficulties
generated by its global properties appear too hard to overcome [14,15,17,18]. 

\section{Conclusion}

QDRK remains the best suited model for 
LSV in UHCR phenomenology.  
Requiring simultaneously the absence of GZK cutoff in the region
$E~\approx ~
10^{20}~eV$~, and that cosmic rays with
$E$ below $\approx ~3.10^{20}~eV$ deposit most of their energy in the
atmosphere, leads [11,14] to:
$10^{-72}~cm^2~<~\alpha ~a^2~<~
10^{-61}~cm^2$~, equivalent to $10^{-20}~<~\alpha ~<~10^{-9}$ for
$a~\approx 10^{-26}~cm$~ ($\approx~10^{21}~GeV$ scale).
Assuming full-strength
LSV forces $a$ to be in the range $10^{-36}~cm~<~a~<~
10^{-30}~cm$. 
Thus, the simplest
version of QDRK naturally fits
with the expected potential
role of Planck-scale dynamics. The same considerations apply to the 
the transition
in the internal structure of a UHE proton. Above $E_{trans}$ , the particle 
looks closer to a Planck-scale object than to the conventional particles of 
current literature. Even a small LSV at the Planck scale can be at the origin of dramatic 
changes in the structure and interaction properties   
of particles at much lower energy scales (above $E_{trans}$).  
More details, using explicitly the operator formalism, can be found in references [17,18] 
as well as in subsequent
papers of the same series ({\it Deformed Lorentz symmetry and High-Energy
Astrophysics}, see arXiv.org ).

\section{References}

\re
1.\ Amelino-Camelia, G.\ 2002a, paper gr-qc/0210063 of arXiv.org
\re
2.\ Amelino-Camelia, G.\ 2002b, paper gr-qc/0207049 of arXiv.org
\re
3.\ Chechin V.A., Vavilov Yu.N.\ 1999, Proc. ICRC 1999, paper HE.2.3.07
\re
4.\ Gonzalez-Mestres L.\ 1995, paper hep-ph/9505117
of arXiv.org
\re
5.\ Gonzalez-Mestres L.\ 1996, paper hep-ph/9610474 of arXiv.org
\re
6.\ Gonzalez-Mestres L.\ 1997a, paper physics/9702026 of arXiv.org
\re
7.\ Gonzalez-Mestres L.\ 1997b, paper physics/9704017 of arXiv.org
\re
8.\ Gonzalez-Mestres L.\ 1997c, paper physics/9705031 of arXiv.org
\re
9.\ Gonzalez-Mestres L.\ 1997d, paper nucl-th/9708028 of arXiv.org
\re
10.\ Gonzalez-Mestres L.\ 1997e, paper physics/9709006 of arXiv.org
\re
11.\ Gonzalez-Mestres, L.\ 1997f, paper physics/9712047 of arXiv.org
\re
12.\ Gonzalez-Mestres, L.\ 1997g, paper physics/9712056 of arXiv.org
\re
13.\ Gonzalez-Mestres L.\ 1999, paper hep-ph/9905430 of arXiv.org
\re
14.\ Gonzalez-Mestres L.\ 2000a, paper physics/0003080 of arXiv.org
\re
15.\ Gonzalez-Mestres L.\ 2000b, paper astro-ph/0011181 of arXiv.org
\re
16.\ Gonzalez-Mestres L.\ 2000c, paper astro-ph/0011182 of arXiv.org
\re
17.\ Gonzalez-Mestres L.\ 2002a, paper hep-th/0208064 of arXiv.org
\re
18.\ Gonzalez-Mestres L.\ 2002b, paper hep-th/0210141 of arXiv.org
\re
19.\ Kirzhnits D.A., Chechin V.A.\ 1971, ZhETF Pis. Red. 4, 261
\re
20.\ Kirzhnits D.A., Chechin V.A.\ 1972, Yad. Fiz. 15, 1051
\re
21.\ Magueijo J. , Smolin L.\ 2001, paper hep-th/0112090 of arXiv.org
\re
22.\ Magueijo, J. and Smolin, L., 2002, paper gr-qc/0207085 of arXiv.org
\endofpaper
\end{document}